\documentclass[12 pt, amsfonts,amsmath, amssymb,color]{article}
\usepackage{authblk}
\def\baselinestretch{1.0}
\usepackage{amsmath,amssymb,amsfonts,latexsym,float,graphics,epsfig}
\usepackage{subfig}
\usepackage{verbatim}
\usepackage{relsize}
\usepackage{hyperref}
\usepackage{graphicx}
\usepackage{bm}
\usepackage{epstopdf}
\usepackage{slashed}
\usepackage{color}
\usepackage{tikz-cd}
\usepackage[utf8]{inputenc}
\usepackage[T1]{fontenc}

\def\be{\begin{equation}}
\def\ee{\end{equation}}
\def\bea{\begin{eqnarray}}
\def\eea{\end{eqnarray}}

\begin{document}

\renewcommand\theequation{\arabic{section}.\arabic{equation}}
\catcode`@=11 \@addtoreset{equation}{section}
\newtheorem{axiom}{Definition}[section]
\newtheorem{theorem}{Theorem}[section]
\newtheorem{axiom2}{Example}[section]
\newtheorem{lem}{Lemma}[section]
\newtheorem{prop}{Proposition}[section]
\newtheorem{cor}{Corollary}[section]

\newcommand{\ben}{\begin{equation*}}
\newcommand{\een}{\end{equation*}}

\let\endtitlepage\relax

\begin{titlepage}
\begin{center}
\renewcommand{\baselinestretch}{1.5}  

\vspace*{-0.5cm}

{\Large {Isochronous oscillator with a singular}}\\
 {\Large {position-dependent mass and its quantization}}
 
\vspace{5mm}
\renewcommand{\baselinestretch}{1}  

\centerline{{\bf Aritra Ghosh$^{*,\#}$\footnote{aritraghosh500@gmail.com, ag34@iitbbs.ac.in}, Bhabani Prasad Mandal$^\dagger$\footnote{bhabani.mandal@gmail.com, bhabani@bhu.ac.in}, Bijan Bagchi$^\ddagger$\footnote{bbagchi123@gmail.com}}}

\vspace{3mm}
\normalsize
\text{$^*$School of Basic Sciences, Indian Institute of Technology Bhubaneswar,}\\
\text{Jatni, Khurda, Odisha 752050, India}\\

\vspace{1.5mm}
\text{$^\#$School of Physics and Astronomy, Rochester Institute of Technology,}\\
\text{Rochester, New York 14623, USA}\\

\vspace{1.5mm}
\text{$^\dagger$Department of Physics, Banaras Hindu University,}\\
\text{Varanasi, Uttar Pradesh 221005, India}\\

\vspace{1.5mm}
\text{$^\ddagger$Department of Applied Mathematics, University of Calcutta,}\\
\text{Kolkata, West Bengal 700009, India}\\

\vspace{2mm}

\begin{abstract}
In this paper, we present an analysis of the equation $\ddot{x} - (1/2x) \dot{x}^2 + 2 \omega^2 x - 1/8x = 0$, where $\omega > 0$ and $x = x(t)$ is a real-valued variable. We first discuss the appearance of this equation from a position-dependent-mass scenario in which the mass profile goes inversely with $x$, admitting a singularity at $x = 0$. The associated potential is also singular at $x = 0$, splitting the real axis into two halves, i.e., $x > 0$ and $x < 0$. The dynamics is exactly solvable for both the branches and so for definiteness, we stick to the $x > 0$ branch. Performing a canonical quantization in the position representation and upon employing the ordering strategy of the kinetic-energy operator due to von Roos, we show that the problem is isospectral to the isotonic oscillator. Thus, the quantum spectrum consists of an infinite number of equispaced levels. The spacing between the energy levels is found to be insensitive to the specific choices of the ambiguity parameters that are employed for ordering the kinetic-energy operator \`a la von Roos.
\end{abstract}
\end{center}
\vspace*{0cm}


\end{titlepage}

\section{Introduction}
In the past few decades, there has been a strong interest in the exploration of the so-called `position-dependent-mass systems' which are mechanical systems with mass functions that depend explicitly on the mechanical coordinate(s) \cite{VR0,VR1,VR2,vonroos,VR3,susy3,VR4,bagchi2,mus1,ganguly,VR5,carinena,cruz,cruz1,Tezcan,cruz2,que,BBnano,koc,Lev,Znojil,cunha,L2quant,mus2,dha,fer,bag5,physicaE,bpm,dirac,singularmass}. In physics, the notion of position-dependent mass derived its motivation originally from condensed-matter physics, especially semiconductor theory \cite{vonroos} (see also, \cite{VR0,VR1,VR2,VR3}). However, the concept of an `effective' mass has found its place in other areas of physics, including nanoscale heterojunctions \cite{BBnano}, compositionally-graded crystals \cite{Gel}, quantum liquids \cite{Qliq}, quantum dots \cite{Ser}, and Helium clusters \cite{Bar}. In particular, there has been some interest in constructing models with singular position-dependent masses that could facilitate a modification in the equilibrium distances in the sites of a crystalline lattice with a defect, thus bringing about clustering near the singularity point \cite{singularmass}. From a theoretical perspective, such systems have been studied from the point of view of exact solvability and coherent states (see for example, \cite{cruz2,physicaE}). Moreover, consequences of a position-dependent mass have also been explored in the context of supersymmetric quantum mechanics (see for example, \cite{susy3,cruz1,bpm}) and the Dirac equation \cite{dirac}. Thus, the study of position-dependent-mass systems, especially finding exactly-solvable ones is an interesting problem in itself and deserves special attention. \\

In this paper, we will be concerned with the following nonlinear equation \cite{nonlocal}:
\begin{equation}\label{GambierEq}
\ddot{x} - \frac{\dot{x}^2}{2x} + 2 \omega^2 x - \frac{1}{8x} = 0, \quad \quad x > 0,
\end{equation} where \(\omega\) is a real and positive parameter and the `overdots' indicate derivatives with respect to time \(t\). It has some intriguing features. The equation (\ref{GambierEq}) can be easily identified to be an equation of the Li\'enard-II class: \(\ddot{x} + f(x) \dot{x}^2 + g(x) = 0\), where \(f(x)\) and \(g(x)\) can be found by comparison with equation (\ref{GambierEq}) above. Such equations emerge naturally from position-dependent-mass Hamiltonians and have been discussed extensively in the literature \cite{L2quant,isolie}. An interesting property (and we will prove this) of the equation (\ref{GambierEq}) is that it supports isochronous oscillations \cite{isolie,CV}, i.e., insensitivity of the time period of oscillation on the energy of the system. A prototypical example of such a system is the harmonic oscillator (ho) while another well-studied example is the isotonic oscillator (io), described by the following potentials, respectively:
\begin{equation}
V_{\rm ho}(x) =  \frac{m \omega^2 x^2}{2}, \quad \quad V_{\rm io}(x) = \frac{m \omega^2 x^2}{2} + \frac{k}{x^2}, \quad \quad m ,  \omega, k > 0. 
\end{equation}
The latter is defined either for \(x > 0\) or \(x < 0\) due to the quadratic singularity at \(x = 0\). In fact, it has been demonstrated that the above two are the only examples of one-dimensional rational potentials that support isochronous oscillations \cite{CV}. Interestingly, both the harmonic and isotonic potentials admit bound states with equispaced spectra \cite{isotonic,isotonic1,isotonic2}, leading to the speculation that isochronous dynamical systems should support equispaced spectra in their quantum counterparts \cite{isospec}. The primary motivation behind this work is to assess the quantum spectrum of the system whose Newtonian equation corresponds to (\ref{GambierEq}). As will be shown, the system (\ref{GambierEq}) is isospectral to the isotonic potential mentioned above, both being defined over the half-line of the real axis which can be taken to be \(x > 0\) without loss of generality (although one may also work with \(x < 0\)). Thus, the goal of this paper is to present an exactly-solvable quantum system which not only admits an equally-spaced spectrum but also carries the interpretation of being a position-dependent-mass system, which, as we mentioned earlier, is also of interest from the point of view of applied physics. For the sake of completeness, we will discuss some classical aspects of the equation (\ref{GambierEq}). \\

With this brief introduction and having explained our motivation, let us present the organization of this paper. In the next section, i.e., in Sec. (\ref{premsec}), we will describe some basic classical aspects of the system (\ref{GambierEq}), including the form of the Hamiltonian function in Sec. (\ref{hamsec}), a proof of isochronicity by employing a nonlocal transformation in Sec. (\ref{isosec}), the exact solution of the Newton's equation (\ref{GambierEq}) in Sec. (\ref{classicalsolsec}), and the fixed points of the system (\ref{GambierEq}) in the \((x,\dot{x})\)-phase space in Sec. (\ref{fixedpoint}). Following this, in Sec. (\ref{quantizationsec}), we will perform a canonical quantization of the system in the position representation and will completely solve the time-independent Schr\"odinger equation. In particular, in Sec. (\ref{effsec}), we will discuss a mapping between the position-dependent-mass Schr\"odinger equation and a constant-mass Schr\"odinger equation in an effective potential and then, utilizing the latter, in Sec. (\ref{specsec}), we will present the spectrum of the system. The paper will be concluded in Sec. (\ref{consec}).

\section{Classical dynamics}\label{premsec}
As mentioned earlier, equation (\ref{GambierEq}) falls into the Li\'enard-II type of second-order differential equations. The Hamiltonian aspects of Li\'enard-II systems have been extensively explored by employing the formalism of the Jacobi last multiplier \cite{L2quant,isolie}. Consider the generic form of a position-dependent-mass Hamiltonian, i.e., 
\begin{equation}\label{PDMH}
H(x,p) = \frac{p^2}{2m(x)} + V(x),
\end{equation} where \(m(x)\) and \(V(x)\) are defined on a particular domain (a subset of \(\mathbb{R}\)) and physically signify the effective mass and the conservative potential. Naturally, the dynamics is conservative in the sense that \(H\) is a constant of motion. The Hamilton's equations turn out to be
\begin{equation}
\dot{x} = \frac{p}{m(x)}, \quad \quad \dot{p} = \frac{p^2}{2m(x)^2} m'(x) - V'(x),
\end{equation} where the `primes' denote derivatives with respect to \(x\). These can be coupled to give 
\begin{equation}\label{PDMsecondorderequationofmotion}
\ddot{x} = - \bigg(\frac{m'(x)}{2m(x)} \bigg) \dot{x}^2 -  \frac{V'(x)}{m(x)}, 
\end{equation} and which takes the form \(\ddot{x} + f(x) \dot{x}^2 + g(x) = 0\), for
\begin{equation}\label{fggenericexp}
f(x) = \frac{m'(x)}{2m(x)}, \quad \quad g(x) = \frac{V'(x)}{m(x)}. 
\end{equation}
It may be remarked that for a case where \(m(x) = 2 V(x)\), both \(f(x)\) and \(g(x)\) coincide. 

\subsection{Hamiltonian corresponding to equation (\ref{GambierEq})}\label{hamsec}
From equation (\ref{GambierEq}), a simple inspection reveals that 
\begin{equation}\label{fggambierexp}
f(x) = -\frac{1}{2x}, \quad \quad g(x) = 2 \omega^2 x - \frac{1}{8x}. 
\end{equation}
Thus, comparing the above-mentioned expressions with (\ref{fggenericexp}) allows one to determine the mass profile \(m(x)\) and the conservative potential \(V(x)\). A particular solution of (\ref{fggenericexp}) for the form of \(f(x)\) and \(g(x)\) given in (\ref{fggambierexp}) is
\begin{equation}\label{mVprofilesGambier}
m(x) = \frac{a}{x}, \quad \quad V(x) = a \bigg(2 \omega^2 x + \frac{1}{8x} \bigg), 
\end{equation} with \(a\) being a real constant which we shall take to be positive, i.e., \(a > 0\). Notice that the mass and the potential are both singular (see also, \cite{singularmass}) at \(x = 0\), therefore splitting the real axis into two disconnected halves, namely, \(x > 0\) and \(x < 0\). If \(a > 0\), the branch \(x < 0\) admits a negative mass (see also, \cite{Lev,Znojil}) while taking \(a < 0\), the branch \(x > 0\) admits a negative mass. Although for the sake of definiteness, we will take both \(x > 0\) and \(a > 0\) for which the position-dependent mass is positive, the solutions (both classical and quantum mechanical) that we will present subsequently are valid for the other branch. The form of the Hamiltonian corresponding to the system (\ref{GambierEq}) explicitly reads
\begin{equation}\label{HGambier}
H = \frac{x p^2}{2 a} + a \bigg(2 \omega^2 x + \frac{1}{8x} \bigg). 
\end{equation}
In Sec. (\ref{quantizationsec}), we will present a canonical quantization of the Hamiltonian (\ref{HGambier}), revealing its intriguing spectrum. However, before that, let us briefly demonstrate the isochronicity of the system (\ref{GambierEq}), following \cite{nonlocal}. 

\subsection{Linearization and isochronicity}\label{isosec}
Prior to exploring quantization, let us take a brief detour and demonstrate that the system (\ref{GambierEq}) supports isochronous oscillations. This will be achieved by linearizing the dynamics, i.e., mapping it to that of the harmonic oscillator. The linearization problem rests upon finding a nonlocal transformation \((x,t) \mapsto (X,T)\) of the form (see \cite{nonlocal} and references therein)
\begin{equation}\label{nonlocal1}
\frac{dX}{X} = F(x) dx + G(x) dt, \quad \quad T = t,
\end{equation} such that \(\frac{d^2 X}{dT^2} + \Omega^2 X = 0\) for some constant \(\Omega > 0\). The functional forms of \(F(x)\) and \(G(x)\) are to be determined from the above-mentioned requirement (should such functions exist). Now, (\ref{nonlocal1}) can be trivially rewritten as
\begin{equation}\label{nonlocal2}
\frac{\dot{X}}{X} = F(x) \dot{x} + G(x), 
\end{equation} where we have used the `overdot' to also indicate differentiation with respect to \(T\), because \(T = t\). A further differentiation with respect to time gives (upon substituting equation (\ref{GambierEq}))
\begin{equation}
\ddot{X} = X \bigg[ \bigg( \frac{dF(x)}{dx} + F(x)^2 + \frac{F(x)}{2x} \bigg)\dot{x}^2 + \bigg( \frac{d G(x)}{d x} + 2 F(x) G(x) \bigg) \dot{x} + \bigg(G(x)^2 - F(x) \bigg(2 \omega^2 x - \frac{1}{8x}\bigg)\bigg) \bigg].
\end{equation}
In order to recover the harmonic oscillator, the following equations should then be satisfied: 
\begin{eqnarray}
\frac{dF(x)}{dx} + F(x)^2 + \frac{F(x)}{2x} &=& 0, \label{firsteq} \\
 \frac{d G(x)}{d x} + 2 F(x) G(x) &=& 0, \label{secondeq} \\
 G(x)^2 - F(x) \bigg(2 \omega^2 x - \frac{1}{8x}\bigg) &=& -\Omega^2. \label{thirdeq}
\end{eqnarray}
A particular solution of (\ref{firsteq}) is \(F(x) = 1/2x \). Plugging this into (\ref{secondeq}) determines \(G(x) \propto 1/x\) which also solves (\ref{thirdeq}) for \(\omega = \Omega\). Thus, the system (\ref{GambierEq}) can be mapped to a harmonic oscillator and since the latter evolves in an isochronous fashion, the equation (\ref{GambierEq}) also supports isochronous oscillations. Therefore, employing a nonlocal transformation, we have demonstrated the isochronicity of the system (\ref{GambierEq}).

\subsection{Solution of classical equation of motion}\label{classicalsolsec}
For the sake of completeness, let us present the analytical solution of the equation (\ref{GambierEq}). Since the Hamiltonian is given by (\ref{HGambier}) which is a constant of motion, i.e., \(H(x,p) = E\) for constant \(E\) on a given trajectory, using \(p = a \dot{x}/x\), we can write
\begin{equation}\label{firstintegral}
\frac{\dot{x}^2}{2} = \frac{E}{a} x - 2 \omega^2 x^2 - \frac{1}{8}.
\end{equation}
It is straightforward to integrate the above-mentioned first-order differential equation with the result being
\begin{equation}\label{classsol}
x(t) = \frac{E}{4\omega^2 a} \bigg[1 + \sqrt{1 - \bigg( \frac{a\omega}{E}\bigg)^2} \sin (2 \omega t + \theta_0) \bigg],
\end{equation} where \(\theta_0\) is an integration constant (so is \(E\) in arriving from (\ref{GambierEq}) to (\ref{firstintegral})). Notice that we have the restriction \(E > a \omega\) and because \(-1 \leq \sin (\cdot) \leq 1\), one finds that \(x(t) > 0\) for all times. Another thing which must be pointed out is that the solution \(x=x(t)\) oscillates with constant frequency \(2\omega\), independent of \(E\). In a sense, therefore, the dynamics is `dual' to that of the half harmonic oscillator with frequency \(\omega\) -- since the latter is restricted to \(x > 0\), the time period of an oscillation is halved while the frequency is doubled. It may be remarked that \(a \rightarrow -a\) takes the classical solution (\ref{classsol}) of the branch \(x > 0\) to \(x < 0\). In a sense, therefore, the transformation \(a \rightarrow -a\) leads to a `conjugated' dynamics if we view the branches \(x > 0\) and \(x < 0\) to be conjugates of each other. Therefore, the dynamics is invariant under the combined transformations \(a \rightarrow -a\) and \(x \rightarrow -x\). In any case, the condition for bounded motion is 
\begin{equation}\label{classcondsol}
\bigg|\frac{E}{a\omega}\bigg| > 1. 
\end{equation}

\subsection{Fixed points}\label{fixedpoint}
Having come this far discussing the classical dynamics dictated by the equation (\ref{GambierEq}), let us comment on the fixed points in the \((x,\dot{x})\)-phase space (note that \(\dot{x}\) is distinctively different from \(p\) due to the position-dependent mass; one is not a mere rescaling of the other) before we move on to assessing the quantum aspects. Denoting \(y = \dot{x}\) (see for example, \cite{fixedbook1,fixedbook2,fixedbook3,bagchidyn1,bagchidyn2}), the system (\ref{GambierEq}) can be expressed as a planar dynamical system, i.e., 
\begin{equation}
\dot{x} = y, \quad \quad \dot{y} = \frac{y^2}{2x} - 2 \omega^2 x + \frac{1}{8x}. 
\end{equation}
The fixed points are obtained straightforwardly by setting \(\dot{x} = 0 = \dot{y}\) at \((x_f,y_f)\), the subscript `\(f\)' denoting a fixed point. This immediately gives
\begin{equation}
y_f = 0, \quad \quad x_f = \pm \frac{1}{4 \omega}. 
\end{equation}
Thus, the fixed points lie on the \(x\)-axis in the \((x,y)\)-phase space. If we restrict our attention to the \(x >0\) branch, there is a single fixed point at \((1/4\omega,0)\). Interestingly, the points on the \(x\)-axis coincide exactly with the minima of the potential \(V(x)\) appearing in (\ref{mVprofilesGambier}); therefore, these are stable fixed points. In other words, the presence of the position-dependent mass does not affect the location of the fixed point(s); we have excluded the point \(x = 0\) from the discussion as both \(m(x)\) and \(V(x)\) are singular at this point and as explained earlier, one must restrict to either \(x > 0\) or \(x < 0\).

\section{Canonical quantization}\label{quantizationsec}
Let us now turn to the canonical quantization of the system (\ref{GambierEq}) with the corresponding Hamiltonian being given by (\ref{HGambier}). While in the position representation, we can use the standard procedure (in units where \(\hbar = 1\)):
\begin{equation}
x \rightarrow \hat{x} = x, \quad \quad p \rightarrow \hat{p} = - i \frac{d}{d x},
\end{equation} therefore leading to \([\hat{x},\hat{p}] = i\), the kinetic-energy term that is proportional to \(x p^2\) requires special care due to the ordering ambiguity that appears because the operators \(\hat{x}\) and \(\hat{p}\) do not commute. There have been various developments concerning the `right' ordering of the kinetic-energy operator of a position-dependent-mass system (see \cite{VR5,physicaE} for detailed discussions) with the most general one being due to von Roos \cite{vonroos} which gives
\begin{eqnarray}
\hat{H}(\hat{x},\hat{p}) = \frac{1}{4} \bigg[ m^\alpha(\hat{x}) \hat{p} m^\beta (\hat{x}) \hat{p} m^\gamma (\hat{x}) + m^\gamma(\hat{x}) \hat{p} m^\beta (\hat{x}) \hat{p} m^\alpha (\hat{x}) \bigg] + V(\hat{x}) , \label{vonroos}
\end{eqnarray} where the `ambiguity parameters' \(\alpha\), \(\beta\), and \(\gamma\) satisfy \(\alpha + \beta + \gamma = -1\). For a generic position-dependent-mass Hamiltonian (\ref{PDMH}), following the above-mentioned ordering strategy, the time-independent Schr\"odinger equation turns out to be \cite{vonroos}
\begin{eqnarray}
- \frac{1}{2 m(x)}  \bigg[\psi'' (x) - \frac{m'(x)}{m(x)} \psi'(x) + \frac{\beta + 1}{2} \bigg( 2 \frac{m'(x)^2}{m(x)^2}  - \frac{m''(x)}{m(x)}\bigg) \psi(x)  \nonumber \\
 + \alpha(\alpha + \beta + 1) \frac{m'(x)^2}{m(x)^2} \psi(x) \bigg] 
+ V(x) \psi(x) = E \psi(x). 
\label{TISEgen}
\end{eqnarray}
For the present system of interest, a great deal of simplification is achieved because \(m(x) = a/x\), leading to
\begin{equation} 
2 \frac{m'(x)^2}{m(x)^2}  = \frac{m''(x)}{m(x)}, \quad \quad \frac{m'(x)}{m(x)} = - \frac{1}{x}.
\end{equation}
Thus, equation (\ref{TISEgen}) simplifies to  
\begin{equation}\label{TISEsimp}
- \frac{1}{2 m(x)}  \bigg[\psi'' (x) + \frac{\psi'(x)}{x} + \frac{\alpha(\alpha + \beta + 1)}{x^2} \psi(x) \bigg] + V(x) \psi(x) = E\psi(x).
\end{equation} 
With equation (\ref{TISEsimp}) being the time-independent Schr\"odinger equation, in what follows, we will describe a transformation to map it to a constant-mass scenario (see also, \cite{cruz,physicaE}). 

\subsection{Effective potential}\label{effsec}
Putting \(m(x) = a/x\), equation (\ref{TISEsimp}) turns out to be
\begin{equation}\label{TISEsimp1}
x \psi''(x) + \psi'(x) + \frac{\alpha(\alpha + \beta + 1)}{x} \psi(x) + 2a (E - V(x)) \psi(x) = 0. 
\end{equation} 
At this stage, it is convenient to introduce the transformation \(x = \eta \xi^2\), where \(0 < \xi < \infty\) and \(\eta\) is some positive constant. This implies that
\begin{equation}\label{TISE1moresimp}
\frac{d^2 \psi(\xi)}{d \xi^2} + \frac{1}{\xi} \frac{d\psi(\xi)}{d\xi} - \frac{\epsilon}{\xi^2} \psi(\xi) + 8 a \eta (E - V(\xi)) \psi(\xi) = 0,
\end{equation}
where \(\epsilon = - 4 \alpha (\alpha + \beta + 1) = 4 \alpha \gamma\). The first-derivative term above can be easily eliminated by defining the similarity transformation 
\begin{equation}\label{linearderiremoval}
\phi(\xi) = \sqrt{\xi} \psi(\xi),
\end{equation} which gives 
\begin{equation}\label{TISE1moresimp1}
\frac{d^2 \phi(\xi)}{d \xi^2}  - \frac{\left(\epsilon - \frac{1}{4}\right)}{\xi^2} \phi(\xi) + 8 a \eta (E - V(\xi)) \phi(\xi) = 0.
\end{equation}
Choosing \(\eta = m_0/4a\) for a constant \(m_0\), one finds that
\begin{equation}\label{TISEunitmassreduced1}
- \frac{1}{2m_0} \frac{d^2 \phi(\xi)}{d\xi^2} + V_{\rm eff} (\xi) \phi(\xi) = E \phi(\xi),
\end{equation}
with
\begin{equation}\label{effpot}
V_{\rm eff} (\xi) = V(\xi) + \frac{\left(\epsilon - \frac{1}{4}\right)}{2 m_0\xi^2}. 
\end{equation}
Thus, we have mapped the Schr\"odinger equation (\ref{TISEsimp1}) of the position-dependent-mass system to that of a particle with constant mass \(m_0\) but moving in an effective potential (\ref{effpot}).

\subsection{Spectrum and wavefunctions}\label{specsec}
\begin{figure}
\begin{center}
\includegraphics[scale=1.2]{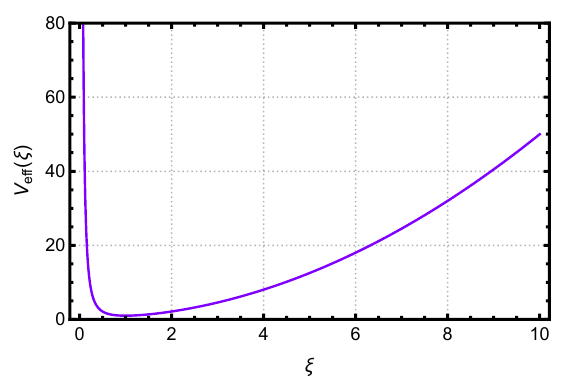}
\caption{Variation of the effective potential \(V_{\rm eff} (\xi)\) for \(m_0 = 1\), \(\omega = 1\), \(a = 1\), and \(\epsilon = 1/4\). This particular choice of \(\epsilon\) arises from choosing the ambiguity parameters \((\alpha,\beta,\gamma)\) to be \(\alpha = \gamma = -1/4\) and \(\beta = -1/2\) \cite{VR5}; other choices \cite{VR0,VR1,VR2,VR3} lead only to minor changes.}
\label{fig}
\end{center}
\end{figure}

\begin{figure}
\begin{center}
\includegraphics[scale=0.9]{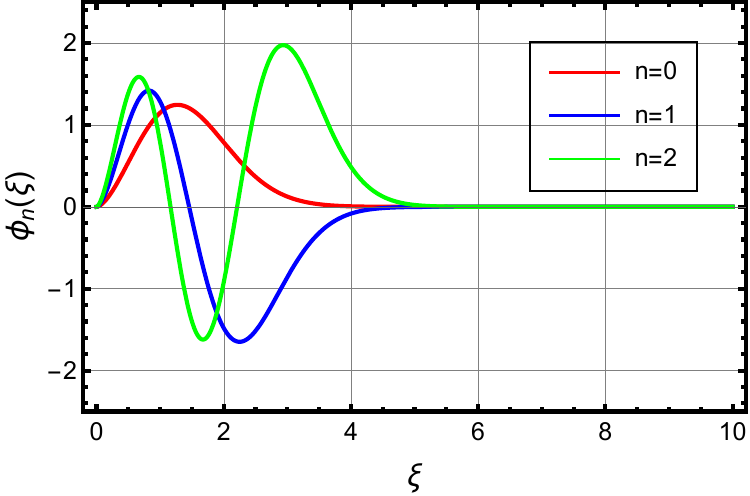}
\caption{Variation of the functions \(\{\phi_n(\xi)\}\) for \(n = 0,1,2\). We have taken \(m_0 = 1\), \(\omega = 1\), \(a = 1\), and \(\epsilon = 1/4\)}
\label{fig2}
\end{center}
\end{figure}

From the form of \(V(x)\) given in (\ref{mVprofilesGambier}), as appropriate for the description of equation (\ref{GambierEq}), and using that \(x = (m_0/4a) \xi^2\), it is easy to verify that the effective potential (\ref{effpot}) explicitly reads
\begin{equation}\label{effpotexplicit}
V_{\rm eff} (\xi) = \frac{m_0 \omega^2 \xi^2}{2} +  \frac{\left(a^2 + \epsilon - \frac{1}{4}\right)}{2 m_0\xi^2}, \quad \quad \xi > 0. 
\end{equation}
This is a remarkable result -- the quantum mechanics of the equation (\ref{GambierEq}) which can be understood as a position-dependent-mass system can be exactly mapped to that of the isotonic oscillator. For the purpose of illustration, the effective potential (\ref{effpotexplicit}) has been plotted in Fig. (\ref{fig}). 
For bounded solutions, one must have
\begin{equation}\label{conditionbound}
a^2 + \epsilon \geq \frac{1}{4}.
\end{equation}
The spectrum and wavefunctions can be easily found because the exact solution for the isotonic potential is well known in the literature for decades \cite{isotonic,isotonic1,isotonic2}. Thus, using the known exact results for the isotonic oscillator \cite{isotonic}, one can write the wavefunctions \(\phi(\xi)\) and the spectrum of the oscillator which turn out to be 
\begin{equation}
\phi_n(\xi) = c_n \xi^\nu e^{- \frac{m_0 \omega \xi^2}{2}} {_1F_1} \bigg(-n; \nu + \frac{1}{2}; m_0 \omega \xi^2\bigg), \quad \quad E_n = \omega \big(2n + 1 + \sqrt{a^2 + \epsilon} \big),
\end{equation} where \(\{c_n\}\) are normalization constants, \(\nu = 0.5 + \sqrt{a^2 +\epsilon}\), \({_1F_1}(~;~;~)\) is the confluent hypergeometric function, and \(n = 0,1,2,\cdots\). Notice that the wavefunctions of the original problem are obtained by inverting (\ref{linearderiremoval}); since \(\xi > 0\), a unique inversion exists. Interestingly, the ground-state energy is 
\begin{equation}
\frac{E_{n=0}}{a \omega} = \sqrt{1 + \frac{\epsilon}{a^2}} + \frac{1}{a} > 0,
\end{equation} due to the condition (\ref{conditionbound}). The functions \(\{\phi_n(\xi)\}\) for \(n = 0,1,2\) have been depicted in Fig. (\ref{fig2}).

\section{Discussion}\label{consec}
In this short paper, we have presented an analysis of the equation \(\ddot{x} - (1/2x) \dot{x}^2 + 2 \omega^2 x - 1/8x = 0\), where \(\omega > 0\) and \(x = x(t)\) is a real-valued variable restricted to the positive half-line. This equation can be obtained from a Hamiltonian that admits a singular position-dependent mass \(m(x) \sim 1/x\) that blows up at \(x = 0\) and falls off for large values of \(x\) (see also, the recent work \cite{singularmass} on singular position-dependent mass). Remarkably, the system supports isochronous oscillations and its spectrum coincides with that of the isotonic oscillator. The wavefunctions can be exactly computed and involve the confluent hypergeometric functions which can in turn be expressed in terms of the associated Laguerre polynomials \cite{handbook}. Thus, the system (\ref{GambierEq}) is an exactly-solvable one, both classically and quantum mechanically. \\

Let us conclude this paper by pointing out the close relationship between the system (\ref{GambierEq}) and the half harmonic oscillator, i.e., a harmonic oscillator with the restriction \(x > 0\). Firstly, the classical dynamics of equation (\ref{GambierEq}) is isochronous with frequency \(2\omega\) as it resembles the harmonic oscillator with frequency \(\omega\) but with dynamics restricted to \(x > 0\) leading to doubling the frequency. Secondly, the spectrum consists of energy levels that are equispaced. The energy spacing is \(\Delta E = E_{n+1} - E_n = 2 \omega\), which coincides with that of a harmonic oscillator of frequency \(\omega\) but with \(x > 0\) -- this is just the half harmonic oscillator! The time-independent Schr\"odinger equation (\ref{TISEsimp1}) can be mapped to that of a constant-mass particle in the isotonic potential. Intriguingly, within the framework of supersymmetric quantum mechanics, the isotonic oscillator is the superpartner of the half harmonic oscillator  \cite{isotonic2,susyiso1} (see also, \cite{davidnew} and references therein for similar discussions on supersymmetric quantum mehanics) and therefore, both of them share the same spectrum (of course, except for the ground-state energy). Interestingly, it may be observed that for the system (\ref{GambierEq}), the energy spacing is independent of the ambiguity parameters \((\alpha,\beta,\gamma)\) employed for ordering the kinetic-energy operator in (\ref{vonroos}) -- these parameters can, however, affect the ground-state energy via the parameter \(\epsilon = 4 \alpha \gamma\).

\section*{Acknowledgements} A.G. and B.B. thank Anindya Ghose-Choudhury, Partha Guha, and Miloslav Znojil for stimulating discussions. A.G. expresses his gratitude to the Ministry of Education, Government of India for financial support in the form of a Prime Minister's Research Fellowship (ID: 1200454). A.G. further thanks the Department of Physics, Banaras Hindu University for hospitality during the final stages of this work. B.P.M. acknowledges the incentive research grant for faculty under the IoE Scheme (IoE/Incentive/2021-22/32253)
 of the Banaras Hindu University.

\section*{Data-availability statement}
 All data supporting the findings of this study are included in the article.

\section*{Conflict of interest}
The authors declare no conflict of interest.

\end{document}